\documentclass[twocolumn,tighten]{aastex631} 
\usepackage{url}

\graphicspath{{./}{figures/}}

\shorttitle{High redshift SPT galaxies}
\shortauthors{Zhen-Yi Cai et al.}

\begin{document}
\def\simlt{\mathrel{\rlap{\lower 3pt\hbox{$\sim$}}\raise 2.0pt\hbox{$<$}}}
\def\simgt{\mathrel{\rlap{\lower 3pt\hbox{$\sim$}}\raise 2.0pt\hbox{$>$}}}

\title{Interpreting the statistical properties of high-$z$ extragalactic sources detected by the South Pole Telescope survey}

\email{zcai@ustc.edu.cn, NegrelloM@cardiff.ac.uk, gianfranco.dezotti@inaf.it}

\author[0000-0002-4223-2198]{Zhen-Yi Cai}
\affiliation{CAS Key Laboratory for Research in Galaxies and Cosmology,
Department  of Astronomy, University of Science and Technology of China, Hefei
230026, China} \affiliation{School of Astronomy and Space Science, University
of Science and Technology of China, Hefei 230026, China}

\author{Mattia Negrello}
\affiliation{School of Physics and Astronomy, Cardiff University, The Parade, Cardiff CF24 3AA, UK}

\author{Gianfranco De Zotti}
\affiliation{INAF, Osservatorio Astronomico di Padova, Vicolo dell'Osservatorio
5, I-35122 Padova, Italy}

\begin{abstract}

The results of the recently published spectroscopically complete survey of dusty star-forming galaxies detected by the South Pole Telescope (SPT) over $2500\,\hbox{deg}^2$ proved to be challenging for galaxy formation models that generally underpredict the observed abundance of high-$z$ galaxies.  In this paper we interpret these results in the light of a physically grounded model for the evolution of spheroidal galaxies. The model accurately reproduces the measured redshift distribution of galaxies without any adjustment of the parameters. The data do not support the indications of an excess of $z>4$ dusty galaxies reported by some analyses of \textit{Herschel} surveys.

\end{abstract}


\section{Introduction}

Ever since the far-IR to mm region opened up to astrophysical investigations it has been a game changer. The InfraRed Astronomy Satellite \citep[IRAS;][]{Neugebauer1984} revealed that in the local universe a substantial fraction ($\simeq 30\%$) of starlight is absorbed and reprocessed by dust, implying that far-IR observations are a key player in extragalactic astrophysics. This became even clearer with Cosmic Background Explorer (COBE) measurements of the Cosmic Infrared Background (CIB) absolute energy spectrum \citep{Puget1996, Fixsen1998} which was found to have approximately the same brightness as the optical background \citep{Dole2006}, implying that a large fraction of starlight was reprocessed by dust and that the reprocessed fraction was higher at high redshifts, as quantitatively predicted by \citet{Franceschini1991}.

The $850\,\mu$m surveys with the Submillimeter Common-User Bolometer Array (SCUBA)  on the James Clerk Maxwell Telescope \citep[JCMT; e.g.,][]{Smail1997, Hughes1998, Barger1998} demonstrated not only that the most active star formation phases of high-$z$ galaxies are heavily dust-enshrouded and therefore largely missed by optical/UV surveys, but also that the abundance of ultra-luminous high-$z$ galaxies is much larger than predicted by the leading merger-driven galaxy formation models of the time  \citep[e.g.,][]{Kaviani2003, Baugh2005} and more consistent with self-regulated galaxy-Active Galactic Nucleus (AGN) co-evolution \citep{Granato2001, Granato2004, Lapi2006, Lapi2011, Lapi2014, Cai2013}.

A new challenge came from searches for sub-mm selected $z\simgt 4$ galaxies using  \textit{Herschel} Spectral and Photometric Imaging Receiver (SPIRE) survey data \citep{Dowell2014, Asboth2016, Ivison2016}; the derived abundances are well above model predictions. The issue is still debated however. \citet{Bethermin2017} argued that due to its limited angular resolution, the SPIRE photometry may be affected by flux boosting due to instrumental noise and confusion (including the contribution from clustering). Some studies \citep{Donevski2018, Duivenvoorden2018} suggested that indeed the discrepancy with models might be accounted for by these effects.

On the other hand, \citet{Cai2020} showed that an excess of high-$z$ galaxies over model predictions had to be expected in the presence of the top-heavy stellar initial mass function (IMF) inferred by \citet{Zhang2018} to account for the low $^{13}$C/$^{18}$O abundance ratio found in four gravitationally lensed sub-mm galaxies at $z\sim 2$--3. A top-heavy IMF was advocated also by \citet{Katz2022} to account for the [OIII]$88\,\mu$m--star formation rate (SFR) and [CII]$158\mu$m--SFR relations observed at $z>6$.

However, firm conclusions on space densities of high-$z$ galaxies were hampered by the uncertainties on source redshifts, which were mostly photometric. The sample of 81 galaxies with full spectroscopic completeness and proper de-boosting \citep{Reuter2020}, drawn from the South Pole Telescope Sunyaev–Zeldovich (SPT--SZ) survey covering $\simeq 2,530\,\hbox{deg}^2$ \citep{Everett2020}, is allowing us to put the analysis on more solid grounds. The theoretical framework is presented in Section~\ref{sect:model}, the completeness of the \citet{Reuter2020} is discussed in Section~\ref{sect:sample}, and in Section~\ref{sect:model_data} the observed redshift distribution is compared to model predictions for different IMFs. The main conclusions are summarized in Section~\ref{sect:conclusions}.

We adopt a flat $\Lambda$CDM cosmology  with parameters derived from Planck CMB power spectra: $H_0 = 67.4\,\hbox{km}\,\hbox{s}^{-1}\, \hbox{Mpc}^{-1}$ and $\Omega_{\rm m} = 0.315$
\citep{PlanckCollaboration2020parameters}.

\begin{figure*}[ht!]
\plotone{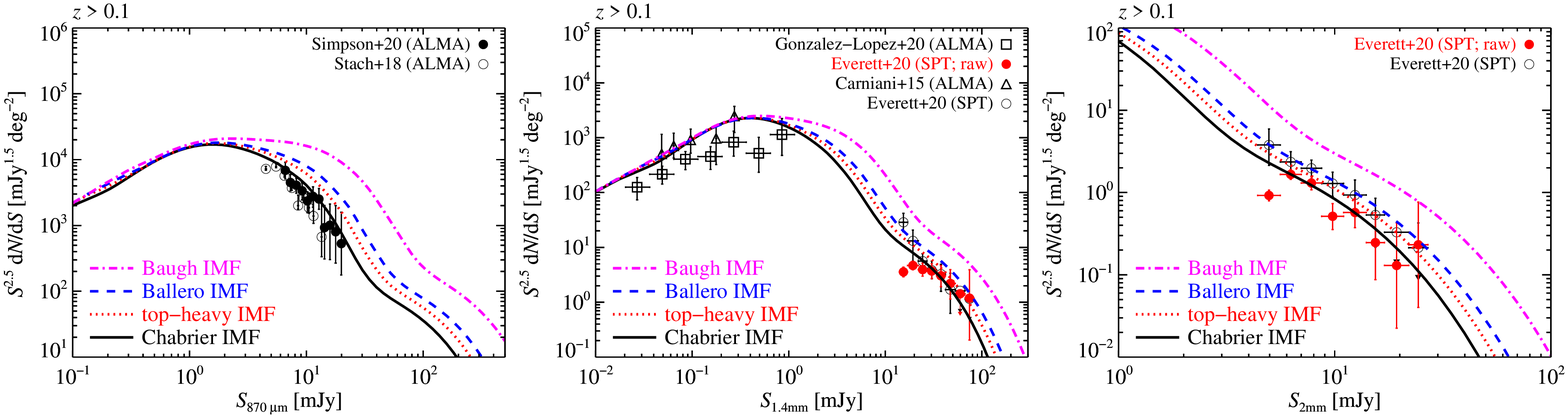}
\caption{Euclidean normalized differential counts at $870\,\mu$m, 1.4\,mm, and 2\,mm: predictions of the \citet{Cai2013} model (solid black lines) compared with observational estimates by \citet{Stach2018} and \citet{Simpson2020} at $870\,\mu$m, and of \citet{Everett2020} at 1.4 and 2\,mm. The latter counts are the ``$z$ cut'' ones, i.e., those obtained removing galaxies at $z\le 0.1$; obviously such galaxies are removed also from model predictions.  The filled red circles show our own counts at 1.4 and 2\,mm derived from the \citet{Everett2020} catalogue, not corrected for incompleteness (see text). At 1.4\,mm we have added the deep counts by \citet{Carniani2015} at 1.3\,mm and by \citet{GonzalezLopez2020} at 1.24\,mm. The flux densities at the latter wavelengths have been scaled to 1.4\,mm using the proto-spheroidal SED by \citet{Cai2013}. The scaling factors are very weakly dependent on redshift for $z<8$; their mean values are 0.80 and 0.65, respectively. The dotted, dashed, and dot-dashed lines refer to top-heavier IMFs specified in the inset (see Section~\ref{sect:model_data}).
\label{fig:counts}}
\end{figure*}


\section{Theoretical Framework}\label{sect:model}

The ages of stellar populations in galaxies demonstrate \citep[see Figure\,10 of][]{Bernardi2010} that at $z\simgt 1.5$, i.e., in the redshift range of interest here, most of the star formation activity is associated to massive proto-spheroidal galaxies, consistent with the downsizing scenario \citep[e.g.,][]{Thomas2010}. Evidence of a morphological transition of dusty galaxies at $z\sim 1.25$ was reported by \citet{LingYan2021}, based on Hubble Space Telescope (HST) images of samples from the \textit{Herschel} Multi-tiered Extragalactic Survey \citep[HerMES;][]{Oliver2012} and from SCUBA2 surveys of the Cosmic Evolution Survey \citep[COSMOS;][]{Scoville2007} field, covering the redshift range $0.5<z<3$. At lower redshifts such galaxies are predominantly disk galaxies, at higher redshifts they are predominantly irregular/interacting systems which are expected to evolve into spheroidal galaxies. Indications of morphological evolution, occurring at $z\sim 1.4$, were pointed out by \citet{Zavala2018a}. These authors argued that data on galaxies detected by their very deep SCUBA2 survey, imaged with the HST, can be interpreted in terms of an evolutionary path whereby galaxies classified by them ``irregular disks'' are the precursors of present day ellipticals.

A physically grounded model for the formation and evolution of these objects was provided by \citet{Cai2013}. The model adopts as the halo formation rate, a function of halo mass and redshift, the positive term of the time derivative of the halo mass function. Such derivative was computed using the analytical approximation by \citet{ShethTormen1999}. The star formation and the growth of the active nucleus (the AGN) are triggered by the first, fast collapse phase of the halo, including major mergers, as highlighted by high resolution $N$-body simulations \citep[e.g.,][]{Wang2011}. The subsequent slow growth of the halo outskirts by minor mergers and diffuse accretion has little effect on the inner part of the potential well, where the visible galaxy resides. Star formation and nuclear activity are governed by in-situ processes described by a set of equations  including gas cooling, condensation into stars, radiation drag, accretion onto the central supermassive black hole,  feedback from supernovae and from the active nucleus. Numerically solving these equations we obtain SFRs and accretion rates as a function of halo mass, formation redshift, and galactic age.

SFRs are converted into total infrared (IR; 8--$1000\,\mu$m) luminosities, $L_{\rm IR}$, using the standard calibration \citep[e.g.,][]{KennicuttEvans2012}. The bolometric luminosity functions of galaxies, of AGNs, and of objects as a whole (galaxy plus AGN) at each $z$ are obtained coupling the luminosity as a function of halo mass with the halo formation rate. Monochromatic luminosity functions of galaxies are derived adopting appropriate spectral energy distributions. The effect of gravitational lensing on observed counts and luminosity functions is also taken into account.

This approach was the only one which successfully predicted the SPT and the \textit{Herschel} Spectral and Photometric Imaging Receiver (SPIRE) counts of strongly lensed galaxies \citep{Vieira2010, Negrello2010}. The \citet{Cai2013} model accurately reproduced a broad variety of multi-frequency data (source counts, redshift distributions, multi-epoch luminosity functions) as reported in the paper itself, as well as later data \citep{Cai2014, Carniani2015, Bonato2014a, Bonato2017, Bonato2019, Gralla2019, DeZotti2019}; see also Figure~\ref{fig:counts}.

\citet{Cai2020} addressed the issue of the abundance of $z>4$ galaxies detected by \textit{Herschel}/SPIRE surveys pointing out the important effect of the stellar IMF.  The interpretation of the controversial excess of $z>4$ dusty galaxies in terms of a top-heavy IMF was however inconclusive, largely due to uncertainties on photometric redshifts. The SPT sample with fully spectroscopic redshifts is thus an important advance. However, to properly compare the observed redshift distribution with model predictions we need to assess the completeness of the \citet{Reuter2020} sample.

\begin{table*}
\caption{SPT sources in the \citet{Everett2020} ``z cut'' sub-sample with $S_{220\,\rm GHz, deb} \ge 20\,$mJy \citep[note that $S_{\rm deb}$ is called $S_{\rm best}$ by][]{Everett2020}  not in the \citet{Reuter2020} sample. Flux densities are in mJy. Photometric redshifts listed in the last column were obtained as described in the text. The errors on $z_{\rm phot}$ are those yielded by the fitting routine. The real uncertainties are much larger and associated to the choice of the SED used for the fit. The four sources in boldface marked with asterisks are probably high-$z$ dusty galaxies (see Section~\ref{sect:sample}).
}
\label{tab:additional}
\begingroup
\footnotesize
\setlength{\tabcolsep}{2pt} 
\hspace*{-2.3cm}
\begin{tabular}{lrrlllcccll}
\hline
  \multicolumn{1}{c}{SPT ID} &
  \multicolumn{1}{c}{RA} &
  \multicolumn{1}{c}{Dec} &
  \multicolumn{1}{c}{$S_{95\,\rm GHz, deb}$} &
  \multicolumn{1}{c}{$S_{150\,\rm GHz, deb}$ } &
  \multicolumn{1}{c}{$S_{220\,\rm GHz, deb}$} &
  \multicolumn{1}{c}{$S_{500\,\mu\rm m}$} &
  \multicolumn{1}{c}{$S_{350\,\mu\rm m}$ } &
  \multicolumn{1}{c}{$S_{250\,\mu\rm m}$ } &
  \multicolumn{1}{c}{Redshift} &
  \multicolumn{1}{c}{Note}  \\
\hline
{\bf J000613-5620.7$^\ast$} &  1.55824 & -56.34621 & $13.2^{+2.1}_{-2.3}$ & $14.5^{+1.4}_{-1.2}$ & $30.4 \pm 3.0$ & & & $47.6\pm 4.3$ & 0.2627 & $z_{\rm phot} \simeq 4.2 \pm 0.2$\\
J015539-5829.1 & 28.91375 & -58.48602 & 19.7 $\pm 2.2$    &\ $11.5^{+1.4}_{-1.3}$ &\ 20.2 $\pm 4.1$ & & & & none & Radio source \\ 
J032538-5247.1 & 51.41249 & -52.78567 &\ $2.1^{+1.7}_{-1.0}$  &\ 5.9 $\pm 1.2$   &\ $20.9^{+4.1}_{-4.3}$ & & & & none & IC\,1933 \\
{\bf J032837-6447.4$^\ast$} & 52.15700 & -64.79109 &\ $1.0^{+0.7}_{-0.3}$  &\ $5.7^{+1.0}_{-0.9}$ &\ 21.8 $\pm 3.5$   & & & & none & $z_{\rm phot} \simeq 1.0 \pm 0.002$ \\
J051445-6449.1 & 78.69093 & -64.81905 &\ $1.5^{+1.4}_{-0.6}$  &\ $6.5^{+1.4}_{-1.0}$ &\ 24.5 $\pm 4.9$   &$98.9\pm5.7$ &$72.6\pm3.8$ & -- & none & PGCC \\
{\bf J201445-4152.0$^\ast$} &303.69122 & -41.86763 &\ $2.2^{+1.9}_{-1.1}$  &\ 7.2 $\pm 1.4$   &\ $22.2^{+5.0}_{-5.1}$ &$115.6\pm5.0$ &$117.8\pm4.1$ &$69.3\pm3.5$& none & $z_{\rm phot} \simeq 3.6 \pm 0.1$\\
{\bf J213230-4537.8$^\ast$} &323.12610 & -45.63134 &\ 7.3 $\pm 2.5$    &\ $5.0^{+1.2}_{-1.0}$ &\ $22.4^{+4.7}_{-5.3}$ & & & $51.9\pm5.4$& 1.332 & $z_{\rm phot} \simeq 3.9 \pm 0.2$\\
J225737-6116.0 &344.40634 & -61.26759 &\ 7.8 $\pm 2.3$    &\ 6.6 $\pm 1.3$   &\ $24.1^{+4.8}_{-5.0}$ & & & & none & Radio source, cirrus \\
J232216-4836.2 &350.56729 & -48.60440 & $11.5^{+2.3}_{-2.1}$  &\ $7.7^{+1.4}_{-1.2}$ &\ 22.3 $\pm 4.9$   & & & & none & Radio source \\
\hline\end{tabular}
\endgroup
\end{table*}

\begin{figure*}
    \centering
    \includegraphics[width=0.32\textwidth]{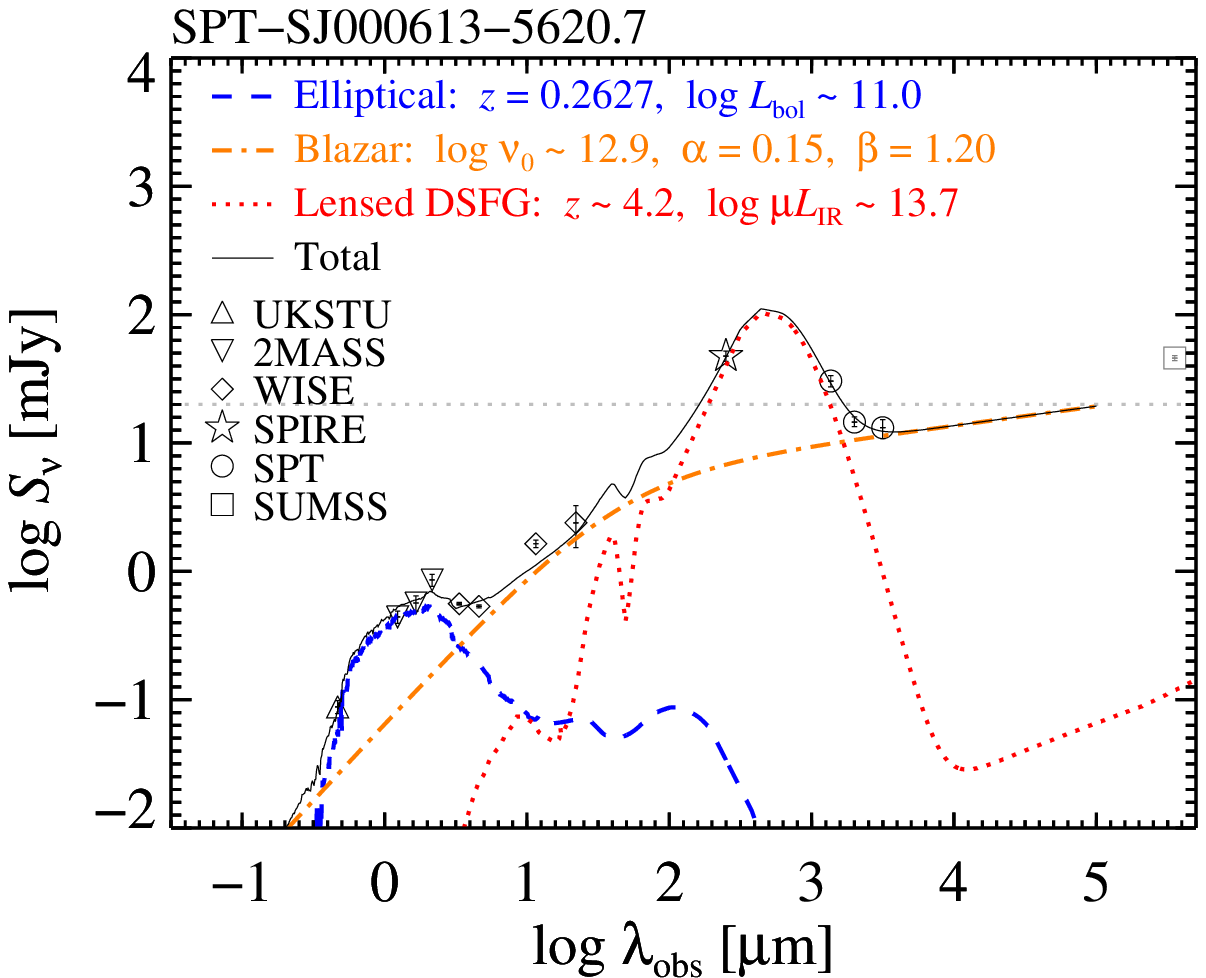}
    \includegraphics[width=0.32\textwidth]{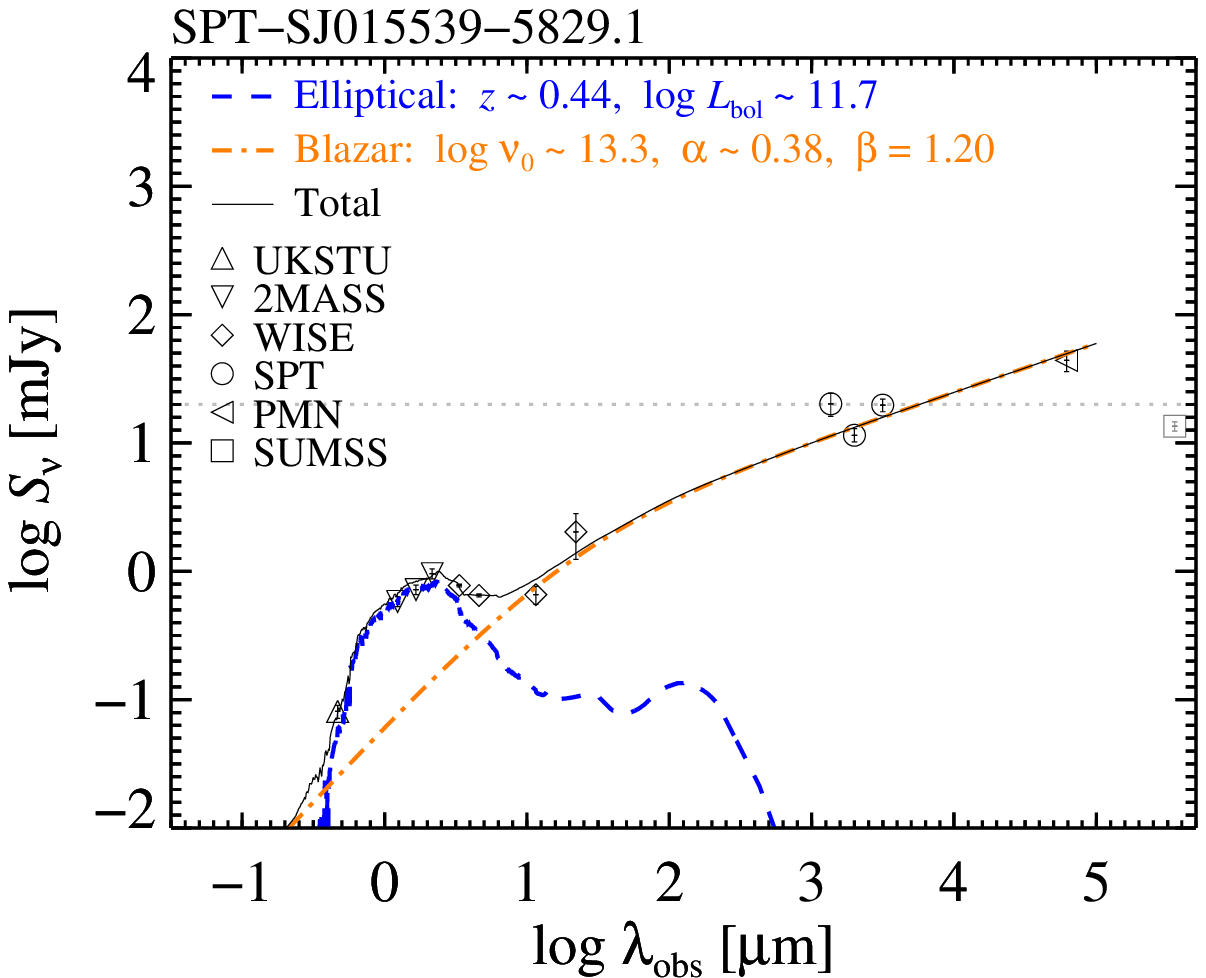}
    \includegraphics[width=0.32\textwidth]{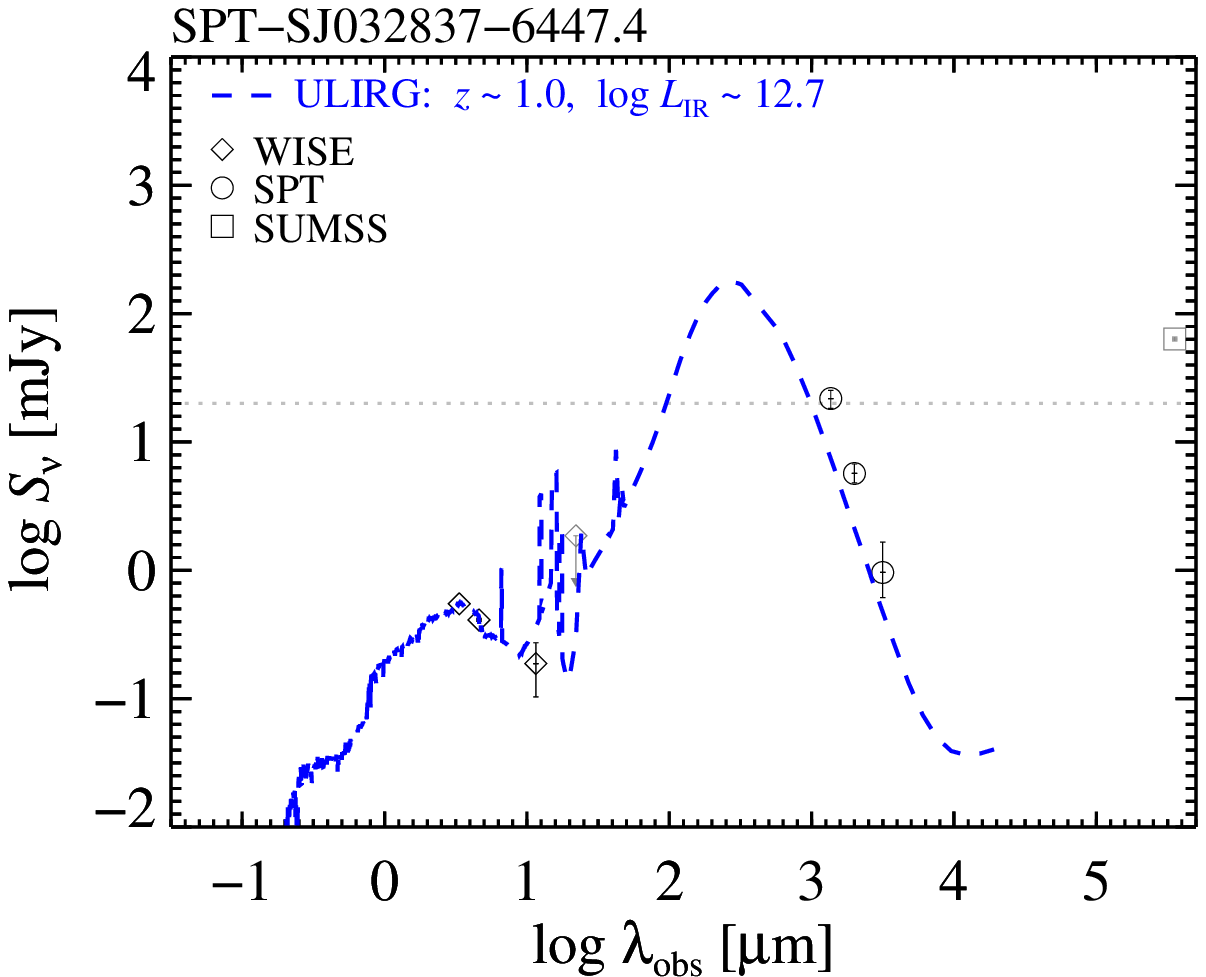}
   \includegraphics[width=0.32\textwidth]{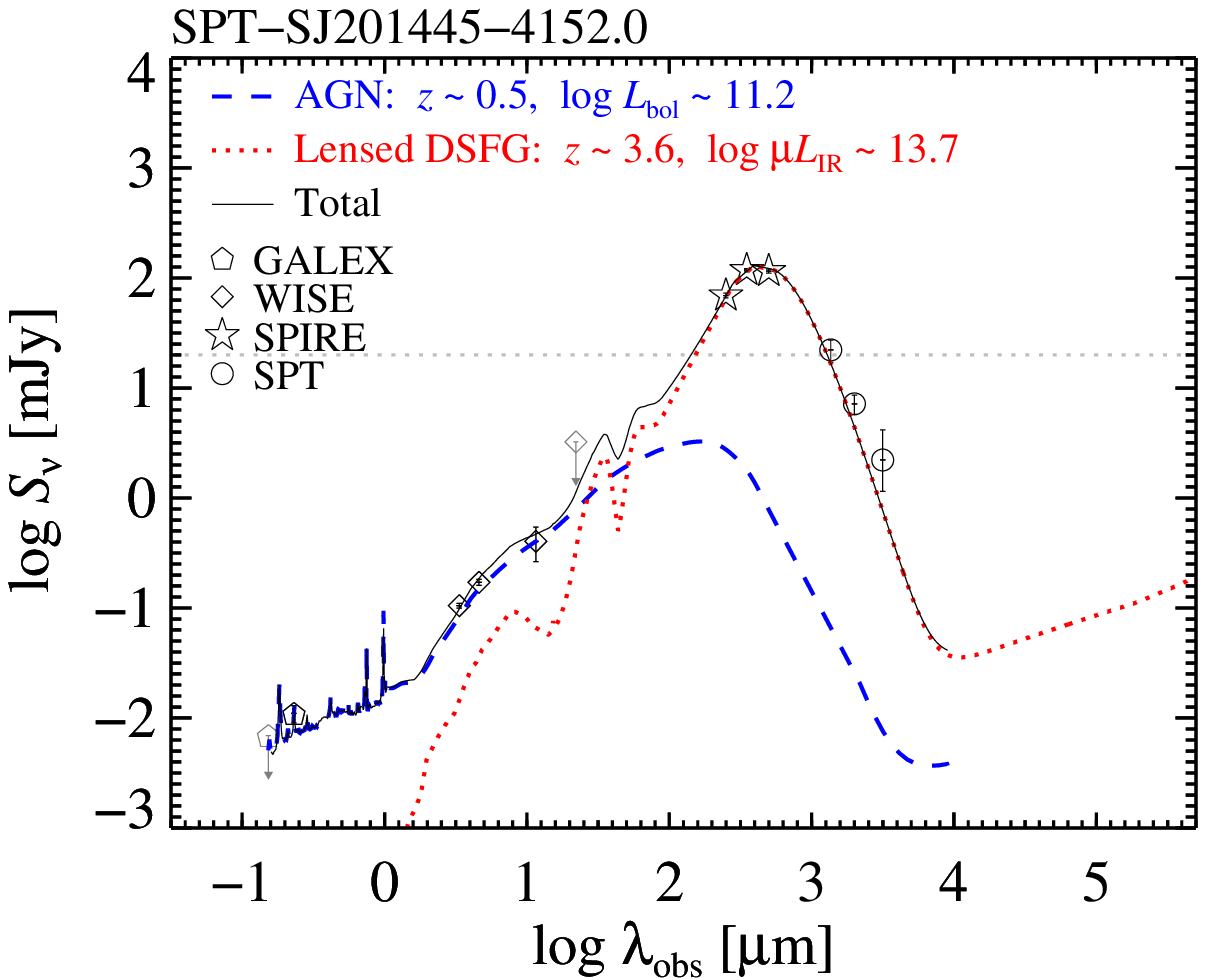}
    \includegraphics[width=0.32\textwidth]{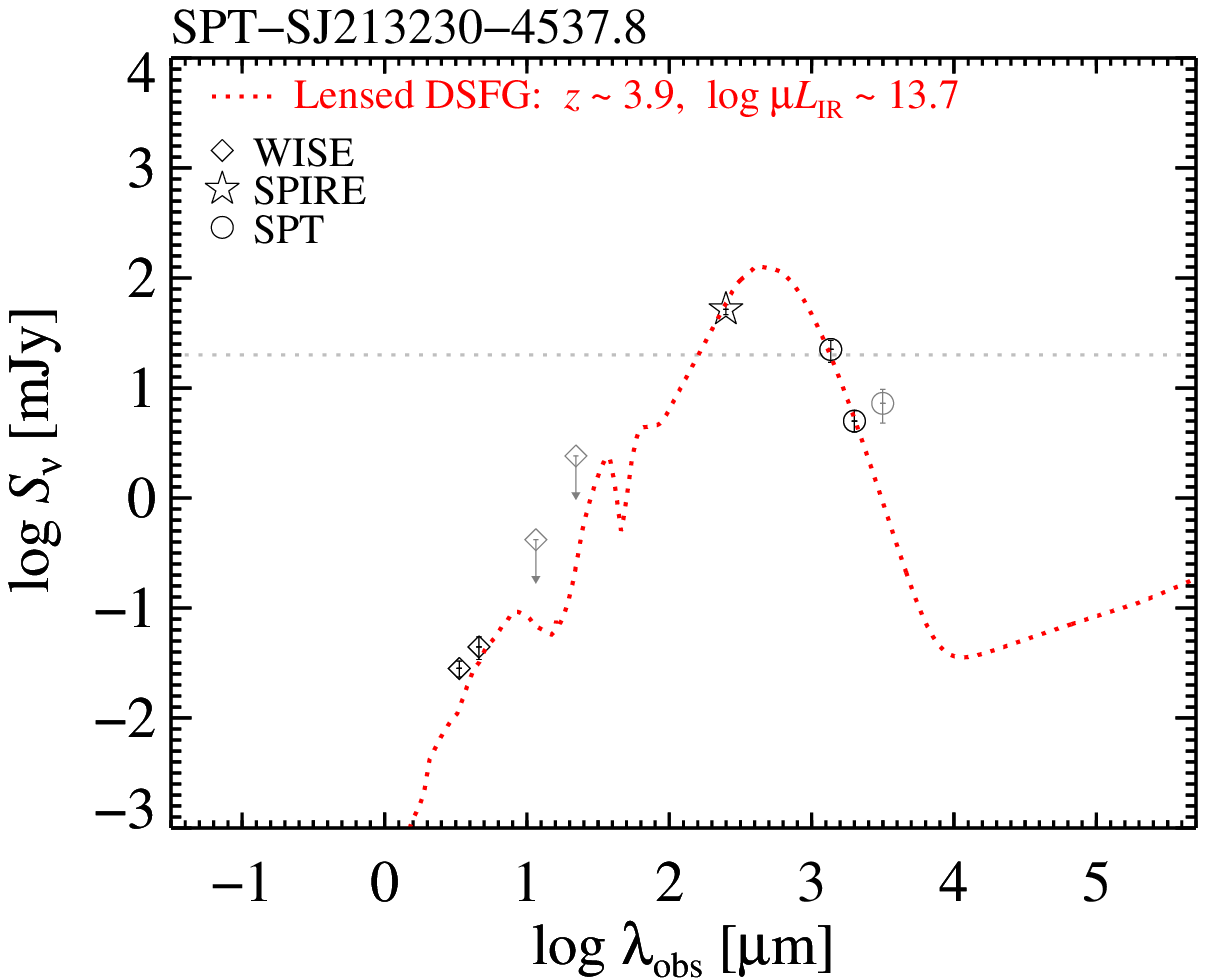}
    \includegraphics[width=0.32\textwidth]{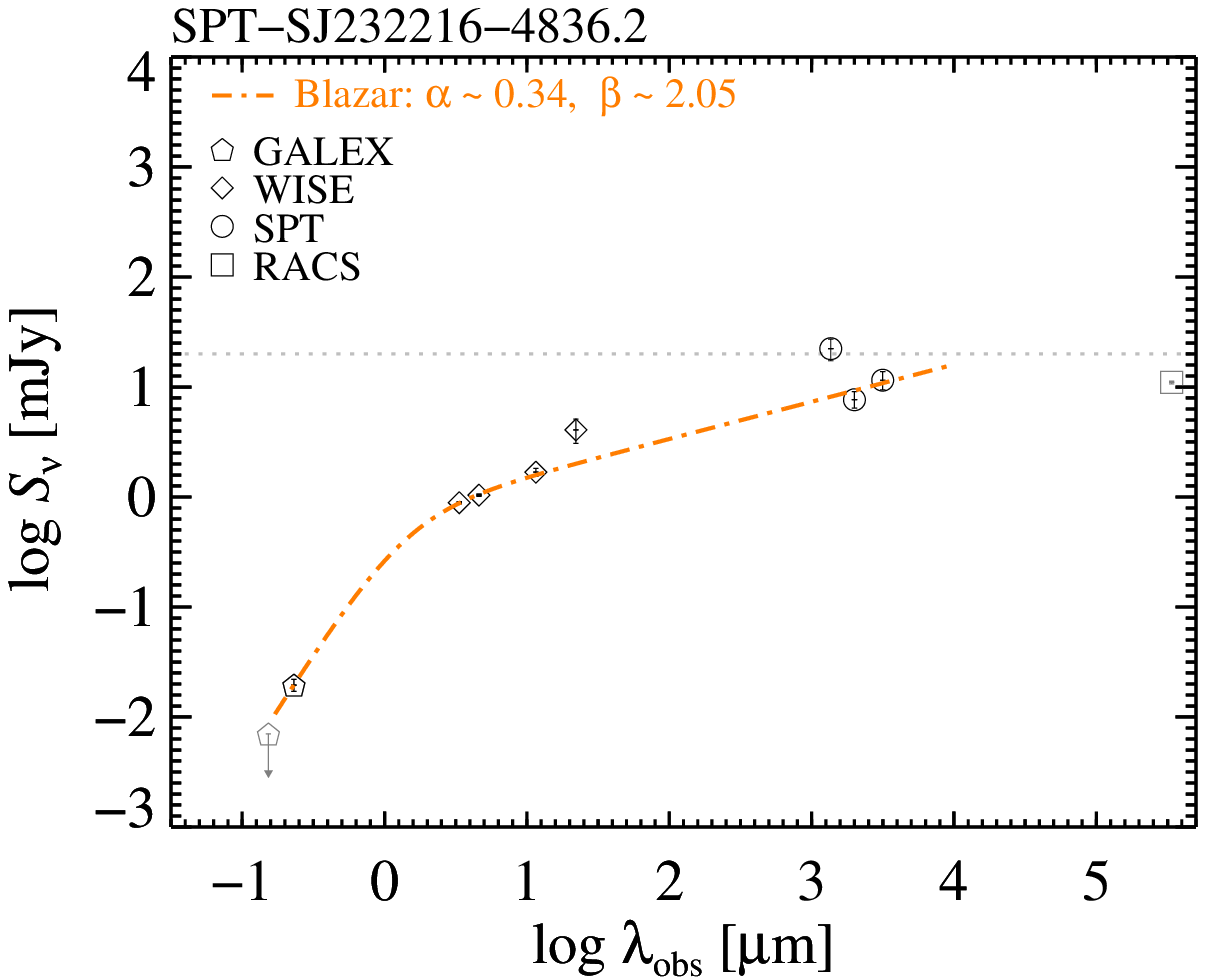}
\caption{
Fits of multi-wavelength data for the six sources in Table~\ref{tab:additional} discussed in the text. Extra data on likely counterparts are from the Galaxy Evolution Explorer \citep[GALEX;][]{Martin2005}, the United Kingdom Schmidt Telescope Unit \citep[UKSTU;][]{Maddox1990}, the Two Micron All Sky Survey \citep[2MASS;][]{Skrutskie2006}, the  Wide-field Infrared Survey Explorer \citep[WISE;][]{Wright2010}, the Parkes-MIT-NRAO (PMN) survey \citep{Wright1994}, the Sydney University Molonglo Sky Survey \citep[SUMSS;][]{Mauch2003}, and the Rapid ASKAP Continuum Survey \citep[RACS;][]{Hale2021}. A single SED or a combination of two/three SEDs, including the \citet{Cai2013} proto-spheroidal SED (red dotted line), a specific SWIRE SED (blue dashed line), and/or a typical blazar SED (orange dot-dashed line), are used to fit the data in black (see text).
}
\label{fig:e20_fit_z_photo}
\end{figure*}

\section{Completeness of the sample}\label{sect:sample}

The \citet{Reuter2020} contains 45 galaxies with deboosted flux density at 1.4\,mm (220\,GHz) $S_{1.4\,\rm mm, deb}\ge 20\,$mJy. The adopted limit is a trade-off between sample size and completeness. The \citet{Reuter2020} sample was generated from a parent sample of $4.5\,\sigma$ detections above a raw flux density of approximately 20.4\,mJy at 1.4\,mm \citep{Everett2020}. As discussed below, the completeness of the \citet{Everett2020} sample above the chosen flux limit is $\simeq 50\%$.

One of the \citet{Reuter2020} sources (SPT2037-65) is not included in the parent catalogue, probably because it lies at the edge of the survey field and was cut in the mask construction process (J. Vieira, personal communication). According to the discussion in \citet{Reuter2020}, none of the 45 galaxies shows indications of multiplicity, i.e., of being a proto-cluster candidate. Also, none of these sources is included in the sample of proto-cluster core candidated by \citet{WangG_2020}.

In the \citet{Everett2020} catalogue there are 9 additional ``dust'' type sources with $S_{1.4\,\rm mm, best}\ge 20\,$mJy and a cut classification ``2'' (Table~\ref{tab:additional})\footnote{\citet{Reuter2020} and \citet{Everett2020} use a slightly different notation for deboosted flux densities, denoted as $S_{\rm deb}$ and $S_{\rm best}$, respectively.}. This flag identifies the ``z cut'' sub-sample which excludes all sources flagged as stars or with cross--matched redshifts $z\le 0.1$  or with angular sizes $\simgt 1'$ (see Section~4.10 of \citealt{Everett2020} for details). Therefore, these 9 sources are expected to be high-$z$ strongly lensed galaxies. One of them (J032538-5247.1), however, can be identified with the local galaxy IC\,1933 and we dropped it.

We searched for the other 8 sources in the \textit{Herschel}/SPIRE point source catalogue\footnote{\url{http://archives.esac.esa.int/hsa/whsa/}} using a search radius of $35''$, the quadratic sum of the $3\,\sigma$ SPT (at 1.4\,mm) and SPIRE astrometric uncertainties, $34''$ \citep[cf. Section~3.7 of][]{Everett2020} and $\simeq 10''$ \citep{Bourne2016}, respectively. We have chosen the FLUX photometry that, according to the SPIRE Point Source Catalog Explanatory Supplement\footnote{\url{archives.esac.esa.int/hsa/legacy/HPDP/SPIRE/SPIRE-P/SPSC/SPIREPointSourceCatalogExplanatorySupplementFull20170203.pdf}} ``has proven to be superior to a number of other common methods used with SPIRE data in terms of reproducibility and photometric accuracy, down to fluxes of 30\,mJy \citep{Pearson2014}''. We retrieved photometric data for 4 galaxies (see Table~\ref{tab:additional}). A source was detected by SPIRE also in the field of J225737$-$6116.0 but, as discussed below, it is most likely associated to a nearby radio source. In fact, the SPIRE flux densities ($46.8\pm3.6$, $52.6\pm3.3$, and $46.5\pm3.2$ at 500, 350, and $250\,\mu$m, respectively) do not match the SPT photometry.

Photometric redshifts were estimated by fitting the \textit{Herschel}/SPIRE and the SPT photometry (except for that at 95\,GHz which may be contaminated by radio emission) with the spectral energy distribution (SED) of high-$z$ proto-spheroidal galaxies by \citet[][see their Figure~\ref{fig:e20_fit_z_photo}]{Cai2013}. The best-fit photo-$z$'s and their 68\% confidence errors, were computed using the routine MPFIT\footnote{\url{https://pages.physics.wisc.edu/~craigm/idl/fitting.html}}, which performs a $\chi^2$ minimization. The results are reported in Table~\ref{tab:additional}.

In the \citet{Everett2020} catalogue, a $z=0.2627$ is ascribed to J000613-5620.7, which is $5''.6$ apart from the SPT position. This redshift was measured for a radio-galaxy, WISEA J000613.47-562042.4. However, the  $250\,\mu\hbox{m}/1.4\,$mm flux density ratio of J000613-5620.7 is indicative of a much higher redshift.  The $z=0.2627$ may belong to the galaxy acting as the lens. The top-left panel of Figure~\ref{fig:e20_fit_z_photo} shows the SPT plus SPIRE photometry together with the data on the radio galaxy reported in the NASA/IPAC Extragalactic Database (NED). The global SED can be interpreted as the sum of a background dusty galaxy at $z\simeq 4.2$ with the radio galaxy. The synchrotron component of the latter has a blazar SED, which, in the considered frequency range, can be represented as \citep{Massardi2022}:
\begin{equation}\label{eq:blazar}
S_\nu=A/\left[\left(\frac{\nu}{\nu_0}\right)^\alpha+\left(\frac{\nu}{\nu_0}\right)^{\beta}\right],
\end{equation}
with $\alpha \simlt 0.5$ and $\beta > 1$. The fit shown has $\alpha = 0.15$, $\beta = 1.20$, and $\log(\nu_0/\hbox{Hz}) \simeq 12.9$. Note that, here and in the following, we do not include in fit the sub-GHz (SUMSS or RACS) flux densities because they are frequently affected by self-absorption or by excesses due to other components.  The blazar dominates the 95\,GHz flux density and contributes substantially also at 150\,GHz. The near-infrared (NIR)-optical excess is interpreted  as due to the blazar host galaxy whose SED is modelled using the Ell2 template taken from the SWIRE  library\footnote{\url{http://www.iasf-milano.inaf.it/~polletta/templates/swire_templates.html}} \citep{Polletta2007} with a bolometric luminosity $\log(L_{\rm bol}/L_\odot) \simeq 11.0$. The sub-mm peak is fitted by the proto-spheroidal template of \citet{Cai2013} at $z\simeq 4.2$; the IR (8--$1000\,\mu$m) luminosity is  $\log(\mu L_{\rm IR}/L_\odot)\simeq 13.7$, $\mu$ being the gravitational magnification.

The source J015539$-$5829.1 may be identified with the radio source PMN J0155$-$5829 (separation of $10.7''$), with a flux density of $44\pm 8\,$mJy at 4.85 GHz, and with the galaxy WISEA J015539.00-582859.9 (separation of $10''$). The global SED has a blazar shape [Equation~(\ref{eq:blazar})] with $\alpha \simeq 0.38$, $\beta = 1.20$, and $\log(\nu_0/\hbox{Hz}) \simeq 13.3$;  therefore we have removed it from our sample. The NIR-optical excess, likely due to the host galaxy, is fitted by a SWIRE Ell2 template at $z\simeq 0.44$ with a bolometric luminosity $\log(L_{\rm bol}/L_\odot) \simeq 11.7$ (see the top-middle panel of Figure~\ref{fig:e20_fit_z_photo}).

J032837-6447.4 is $\simeq 10''$ away from the galaxy WISEA J032839.00-644728.6. If this galaxy can be identified with the SPT source, a fit using the SWIRE Spi4 SED yields a photometric redshift $z\simeq 1.0$ (see the top-right panel of Figure~\ref{fig:e20_fit_z_photo}) and $\log(L_{\rm IR}/L_\odot)\simeq 12.7$, qualifying it as an ultra-luminous IR galaxy (ULIRG). 

As for J051445$-$6449.1, the SPIRE catalogue contains a $500\,\mu$m detection at $25''.3$ from the SPT position. There is also a $350\,\mu$m detection but $31''$ away from the one at $500\,\mu$m. This suggests that J051445$-$6449.1 is extended with emissions at the two wavelengths peaking at different positions. In fact, this source may be identified with a \textit{Planck} Galactic Cold Clump \citep[PGCC;][]{PlanckCollaboration2016cold_cores}, although we caution that  several PGCCs turned out to be high-$z$ strongly lensed galaxies \citep{Trombetti2021}. The extendness hypothesis is confirmed by the lack of detection by Large Apex Bolometer Camera (LABOCA) observations at $870\,\mu$m with an angular resolution  (full-width at half-maximum, FWHM) of $19''.7$ and median rms sensitivity of 8\,mJy \citep[][J. Vieira, private communication]{Greve2012}. Therefore this source was not considered further.

J201445$-$4152.0 was detected in all three SPIRE bands (nominal separations from the SPT position in the range $10''.2$--$11''.7$, depending on the SPIRE band). The SPT and SPIRE data are reasonably well fitted by the \citet{Cai2013} proto-spheroidal SED at $z\simeq 3.6$, with $\log(\mu L_{\rm IR}/L_\odot)\simeq 13.7$. It is $11''.5$ away from WISEA J201446.90-415207.3, which may be the lens. Its SED is fitted by the SWIRE TQSO1 template at $z \simeq 0.5$ with $\log(L_{\rm bol}/L_\odot) \simeq 11.2$ (see the bottom-left panel of Figure~\ref{fig:e20_fit_z_photo}).

For J213230-4537.8 we found a SPIRE detection at $250\,\mu$m (separation of $21''$). Our fit of the SPT photometry at $> 95\,$GHz and of the SPIRE photometry  yielded $z\simeq 3.9$ and $\log(\mu L_{\rm IR}/L_\odot)\simeq 13.7$; the fitting SED is fully consistent with the WISE flux densities of WISEA J213231.44-453749.0 (separation of $13''$ from the nominal SPT position), which is the likely identification (see the bottom-middle panel of Figure~\ref{fig:e20_fit_z_photo}). This galaxy is $37''.2$ away from WISEA J213228.32-453738.2, which may be identified with the blazar AT20G J213227-453740 (the separation among the nominal positions of the two objects is $2''.9$). Thus the blazar lies within the FWHM of the 95\,GHz beam ($\hbox{FWHM}=1'.7$) and can therefore account for the flux density excess at this frequency (it has a flux density of $73\pm 4\,$mJy at 20 GHz). Its continuum spectrum, however, must drop rapidly above 20 GHz (the SPT flux density at 95\,GHz is $7.3 \pm 2.5\,$mJy) and its observed flux density is substantially attenuated at the higher SPT frequencies because of the lower FWHM's; it is thus likely that the blazar contribution at these frequencies is small.

J225737-6116.0 is $17''.9$ away from the radio source SUMSS\,J225739-611606 which is well within the SPT beam at 95\,GHz. The radio source has a flux density of $143.01\pm 1.88\,$mJy at 0.863\,GHz and, if it is flat-spectrum, it may dominate at $95\,$GHz where the data show an excess over a dusty galaxy SED. Subtracting its contribution at 220\,GHz, the flux density may drop below the adopted threshold of 20\,mJy. There are detections at all three SPIRE wavelengths with separations of $\simeq 6''$ from the radio source (which is therefore the likely counterpart to the \textit{Herschel} source) and at $20''$--$23''$ from the SPT position. The SPT source lies in a cirrus region and was not detected with LABOCA at $870\,\mu$m (J. Vieira, private communication), implying that it is  not a dusty galaxy and will not be considered further.

J232216-4836.2 is $2''.2$ away from the radio source RACS-DR1 J232215.9-483615 \citep{McConnell2020, Hale2021} with a flux density of $11.0\pm 0.3\,$mJy/beam at 0.855\,GHz. In turn, the RACS source is $1''.4$ apart from WISEA J232215.92-483616.2. Identifying the 3 sources we get the SED shown in the bottom-right panel of Figure~\ref{fig:e20_fit_z_photo}, indicating that this source is a blazar. It was therefore excluded from our sample.

In conclusion we decided to keep J000613-5620.7, J032837-6447.4, J201445-4152.0, and J213230-4537.8, while J015539-5829.1, J051445-6449.1, J225737-6116.0, and J232216-4836.2 were dropped.

We have included the 3 additional likely strongly lensed sources with photometric redshift estimates in the redshift distribution for $S_{1.4\,\rm mm, best}\ge 20\,$mJy (J032837$-$6447.4 is probably an unlensed ULIRG).  This brings the completeness of the \citet{Reuter2020} sample at the level of the \citet{Everett2020} sample. The 3 sources have redshifts in the range 3.6 -- 4.2 and fill the dip shown by the \citet{Reuter2020} sample.

The completeness of the latter is discussed by \citet{Everett2020}. The area
was divided into 19 contiguous fields, observed independently. The flux density
corresponding to 95\% completeness varies from field to field. At 220\,GHz it
ranges from 16.19 to 29.70\,mJy with a median of 26.83\,mJy. The completeness
as a function of flux density, $f_{\rm compl}(S)$, was calculated by
\citet{Everett2020} adding random locations to 100 simulated sources at fixed
flux density values and applying their source extraction algorithm. The process
was repeated at a few flux density levels spread over a broad range; $f_{\rm
compl}(S)$ was computed as the ratio between the numbers of recovered and input
sources. An error function was fitted to the results and adopted as a model for
$f_{\rm compl}(S)$.

\begin{figure}[ht!]
\plotone{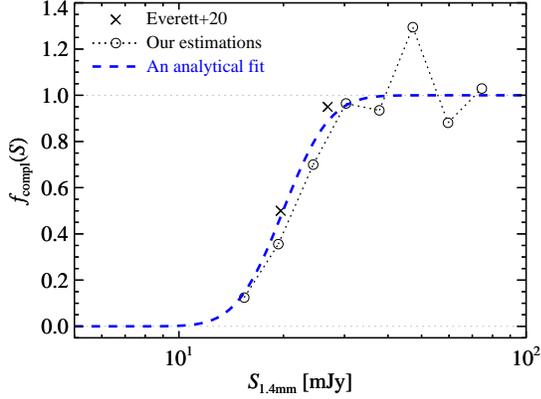}
\caption{Completeness of the ``$z$ cut'' sample at 1.4\,mm (220\,GHz) as a function of flux density, computed as described in the text. The open circles connected by the dotted line show the estimates for each flux density bin. The dashed blue line is the best fit function, $f_{\rm compl}(S)$ [Equation~(\ref{eq:completeness})]. The $\mathbf{\times}$ signs correspond to the median flux densities at the 50\% and 95\% completeness levels across all fields, according to Table\,3 of \citet{Everett2020}, 19.65 and 26.83\,mJy, respectively.
\label{fig:completeness}}
\end{figure}

Since the model $f_{\rm compl}(S)$ is not given by \citet{Everett2020}, we
have estimated it by comparing the counts of dusty galaxies in the ``z cut''
sub-sample derived directly from the catalogue with those, corrected for
incompleteness, reported by \citet{Everett2020}. The uncorrected counts at
220\,GHz (1.4\,mm) and 150\,GHz (2\,mm) are shown by the filled red circles in
the central and right panels of Figure\,\ref{fig:counts}, respectively; the
counts of ``z cut'' sources by \citet{Everett2020} for the same flux density
bins are shown by the open circles. Our estimates of $f_{\rm compl}(S)$ at 1.4\,mm are
shown in Figure~\ref{fig:completeness} (open circles connected by the dotted
broken line). The best fit model in terms of the error function, represented by the
dashed blue line, writes:
\begin{equation}
f_{\rm compl}(S)= 0.5 \times \{ 1 + \mbox{erf}[ (\log S - 1.3) / 0.15] \}.
	\label{eq:completeness}
\end{equation}
The redshift distribution of sources with $S_{1.4\,\rm mm, best}\ge 20\,$mJy
was then computed assigning to each source, including those with just a
photometric redshift estimate, the weight $1/f_{\rm compl}(S)$ corresponding to
its flux density.

\begin{figure}[ht!]
\plotone{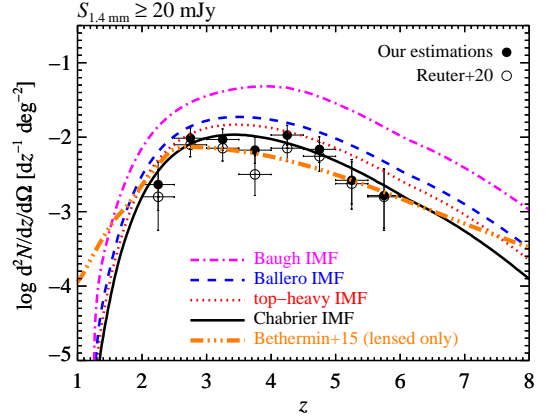}
\caption{Redshift distribution of SPT--SZ sources brighter than $S_{220\,\rm GHz, deb}=20\,$mJy, corrected for incompleteness (filled black circles). The open circles show the redshift distribution of the 45 galaxies with $S_{220\,\rm GHz, deb} \geqslant 20\,$mJy in the \citet{Reuter2020} sample, without any correction for incompleteness. The solid black line shows the prediction of the baseline model by \citet{Cai2013}, adopting the \citet{Chabrier2003} IMF. The other lines show the effect of adopting the top-heavier IMFs specified in the inset, following \citet{Cai2020}. The redshift distribution predicted by the \citet{Bethermin2015} model is also plotted for comparison (dot-dashed orange line).
\label{fig:zdistr}}
\end{figure}

\begin{deluxetable}{ccccc}
\tablecaption{The redshift distribution of SPT sources with $S_{1.4\,\rm mm, best}\ge 20\,$mJy (see also Figure~\ref{fig:zdistr}). \label{tab:rd_S20mJy}}
\tablewidth{0pt}
\tablehead{
\colhead{$z$} & \colhead{$N_{\rm R20}$} & \colhead{$\frac{d^2N_{\rm R20}}{dzd\Omega}$} & \colhead{$N_{\rm our}$} & \colhead{$\frac{d^2N_{\rm our}}{dzd\Omega}$} \\
\colhead{} & \colhead{} & \colhead{(10$^{-3}$ d$z^{-1}$ deg$^{-2}$)} & \colhead{} & \colhead{(10$^{-3}$ d$z^{-1}$ deg$^{-2}$)}
}
\decimalcolnumbers
\startdata
2.25 & $2^{+2.6}_{-1.3}$  & $1.6^{+2.1}_{-1.0}$ & $2.9^{+2.9}_{-1.6}$  & $2.3^{+2.3}_{-1.3}$ \\
2.75 & $10^{+4.3}_{-3.1}$ & $7.9^{+3.4}_{-2.5}$ & $12.3^{+4.6}_{-3.5}$ & $9.7^{+3.6}_{-2.7}$ \\
3.25 & $9^{+4.1}_{-2.9}$  & $7.1^{+3.2}_{-2.3}$ & $11.8^{+4.5}_{-3.4}$ & $9.3^{+3.6}_{-2.7}$ \\
3.75 & $4^{+3.2}_{-1.9}$  & $3.2^{+2.5}_{-1.5}$ & $8.5^{+4.0}_{-2.9}$  & $6.7^{+3.2}_{-2.3}$ \\
4.25 & $9^{+4.1}_{-2.9}$  & $7.1^{+3.2}_{-2.3}$ & $13.5^{+4.8}_{-3.6}$ & $10.7^{+3.8}_{-2.9}$ \\
4.75 & $7^{+3.8}_{-2.6}$  & $5.5^{+3.0}_{-2.0}$ & $8.7^{+4.1}_{-2.9}$  & $6.9^{+3.2}_{-2.3}$ \\
5.25 & $3^{+2.9}_{-1.6}$  & $2.4^{+2.3}_{-1.3}$ & $3.3^{+3.0}_{-1.7}$  & $2.6^{+2.4}_{-1.4}$ \\
5.75 & $2^{+2.6}_{-1.3}$  & $1.6^{+2.1}_{-1.0}$ & $2.1^{+2.7}_{-1.3}$  & $1.6^{+2.1}_{-1.0}$ \\
\enddata
\tablecomments{We adopted a bin size $\Delta z= 0.5$. The total area of the SPT survey is of $2530\,\hbox{deg}^2$. $N_{R20}$ and $N_{\rm our}$ are the numbers of sources within each redshift bin from the \citet{Reuter2020} raw sample and from our updated sample corrected for incompleteness, respectively. The uncertainties are $1\,\sigma$ Poisson errors computed following \citet{Gehrels1986}.}
\end{deluxetable}

\section{Model versus data}\label{sect:model_data}

In Figure~\ref{fig:zdistr}, the redshift distribution of sources with
$S_{1.4\,\rm mm, best}\ge 20\,$mJy, tabulated in Table~\ref{tab:rd_S20mJy}, is compared with predictions by \citet{Cai2020} for different choices of the IMF. The predictions of the phenomenological model by \citet[][lensed galaxies only]{Bethermin2015}, extracted from Figure~10 of \citet{Reuter2020}, are also shown for comparison. \citet{Bethermin2015} reported the redshift distribution of lensed galaxies with $S_{1.4\,\rm mm}> 20\,$mJy yielded by the phenomenological model by \citet{Bethermin2012}. As illustrated by Figure~\ref{fig:zdistr} the agreement with the data is reasonably good, although, as  pointed out by \citet{Reuter2020}, the predicted distribution peaks at a redshift somewhat lower than is observed.

\citet{Reuter2020} also compare their results with models by \citet{Hayward2013},  \citet{Lagos2019}, and \citet{Lovell2021}. However the comparisons are tricky because these models do not include strong lensing. Published predictions must be extrapolated in frequency and in flux density in a complicated way since the flux density ratios depend on redshift. So the extrapolations cannot be accurate.

Figure\,\ref{fig:zdistr} shows that the baseline model by \citet{Cai2013}, adopting an universal Chabrier IMF, reproduces quite well the observed redshift distribution which is, however, also consistent with the ``top-heavy'' IMF proposed by \citet{Zhang2018}. The ``Ballero'' \citep{Ballero2007}  IMF somewhat over-predicts the redshift distribution while the \citet{Baugh2005} IMF yields far too many high-$z$ galaxies, as already found by \citet{Cai2020}.  In terms of $dN/d\log m$, $m$ being the stellar mass, the ``top-heavy'' and the ``Ballero''  IMFs have power-law slopes above $0.5\,M_\odot$ of $-1.1$ and of $-0.95$, respectively; below $0.5\,M_\odot$ both have a slope of $-0.3$. The \citet{Chabrier2003} IMF has a knee at $1\,M_\odot$ and power-law indices of $-0.4$ and $-1.35$ below and above that mass. The \citet{Baugh2005} IMF is $dN/d\log m= \hbox{constant}$. Although our results are consistent with a universal \citet{Chabrier2003} IMF, the degeneracies of the model leave room for improvements of the fits with the \citet{Zhang2018} IMF or even the \citet{Ballero2007} IMF, by adjusting the model parameters. On the other hand, we cannot get a reasonable fit with the \citet{Baugh2005} IMF.

We conclude that the \citet{Reuter2020} data tighten the constraints on the IMF at high $z$. However, a much larger sample is necessary to settle the issue on the ``universality'' of the Chabrier IMF. An important step forward will be possible with the completion of the ongoing effort to obtain spectroscopic redshifts of bright ($S_{500\,\mu\rm m}> 80\,$mJy) \textit{Herschel}--selected galaxies with photometric redshift $z_{\rm phot}>2$. The sample of such \textit{Herschel} Bright Sources (HerBS), detected by the \textit{Herschel} Astrophysical Terahertz Large Area Survey \citep[H-ATLAS;][]{Eales2010} covering an area of $616.4\,\hbox{deg}^2$, contains 209 galaxies \citep{Bakx2018}. Robust spectroscopic redshifts for 77\% of the HerBS sample have already been acquired \citep{Neri2020, Urquhart2021} using the NOEMA (NOrthern Extended Millimeter Array) and the ACA (Atacama Compact Array) facilities. The recently completed NOEMA large program z-Gal (PI: P. Cox) has yielded  spectroscopic redshifts for 125 galaxies with $S_{500\,\mu\rm m}> 80\,$mJy including also galaxies in HerMES \citep[\textit{Herschel} Multi-tiered Extragalactic Survey;][]{Oliver2012}  fields in the Northern sky and in the equatorial region. The results are not public yet.

\section{Conclusions}\label{sect:conclusions}

We have exploited the almost complete spectroscopic coverage of SPT galaxies with deboosted flux density at 1.4\,mm  $S_{1.4\,\rm mm, deb}\ge 20\,$mJy to investigate their redshift distribution, with particular attention to the controversial excess over model prediction  of $z>4$ galaxies. Such excess might indicate a top-heavier IMF in proto-spheroidal galaxies that dominate the star-formation activity at high $z$ \citep{Cai2020}. Observational evidences and theoretical arguments in this directions have indeed been put forward \citep{Chiosi1998, Zhang2018, Romano2019, Katz2022} but the issue is still open and may have important implications for the understanding of galaxy evolution.

After applying careful corrections for incompleteness of the \citet{Reuter2020} sample and of the \citet{Everett2020} parent sample we found that the redshift distribution is accounted for quite well both by the physical model by \citet{Cai2013}, adopting an universal Chabrier IMF and by the phenomenological model by \citet{Bethermin2015}, although the peak of the distribution predicted by the latter occurs at a redshift somewhat lower than is observed.

The data tighten the constraints on the high-$z$ IMF, compared to the data discussed by \citet{Cai2020}. While they are consistent with the moderately top-heavy IMF proposed by \citet{Zhang2018}, they are in tension with the ``Ballero'' IMF that fitted the earlier data and are strongly inconsistent with the \citet{Baugh2005} IMF.  The much larger sample that will be provided by the ongoing effort to obtain spectroscopic redshifts of bright ($S_{500\,\mu\rm m}> 80\,$mJy) \textit{Herschel}--selected galaxies with photometric redshift $z_{\rm phot}>2$ will allow an important step forward towards settling the issue on the universality of the Chabrier IMF.

\section*{Acknowledgements} We are grateful to the referee for useful comments and to Joaquin Vieira for important information on SPT data.
Z.Y.C. is supported by the National Science Foundation of China (grant Nos. 11873045, 11890693, and 12033006) and the USTC Research Funds of the Double First-Class Initiative (grant No. YD2030002009). This research has made use of the \citet{NED}, which is operated by the Jet Propulsion Laboratory, California Institute of Technology, under contract with the National Aeronautics and Space Administration. We also made use of the Herschel science archive (\url{http://archives.esac.esa.int/hsa/whsa/}), of data products from the Two Micron All Sky Survey (2MASS), from the Wide-field Infrared Survey Explorer (WISE) and from the Galaxy Evolution Explorer (GALEX). The 2MASS is a joint project of the University of Massachusetts and the Infrared Processing and Analysis Center/California Institute of Technology, funded by the National Aeronautics and Space Administration and the National Science Foundation. The WISE is a joint project of the University of California, Los Angeles, and the Jet Propulsion Laboratory/California Institute of Technology, funded by the National Aeronautics and Space Administration. The GALEX satellite was a NASA mission led by the California Institute of Technology.

\bibliographystyle{aasjournal}
\bibliography{highz_submm_galaxies}

\end{document}